\documentclass[twocolumn,astrosymb,trackchanges]{aastex631}

\usepackage{soul}

\begin{document}

\title{Star-Planet Interaction at radio wavelengths in YZ\,Ceti:\\
Inferring planetary magnetic field}





\author[0000-0002-1216-7831]{Corrado Trigilio}
\affiliation{INAF-Osservatorio Astrofisico di Catania, Via Santa Sofia 78, I95123 Catania, Italy}

\author[0000-0002-1741-6286]{Ayan Biswas}
\affiliation{Department of Physics, Engineering Physics \& Astronomy, Queen’s University, Kingston, Ontario K7L 3N6, Canada}
\affiliation{Department of Physics and Space Science, Royal Military College of Canada, PO Box 17000, Station Forces, Kingston, ON K7K 7B4, Canada}
\affiliation{National Centre for Radio Astrophysics, Tata Institute of Fundamental Research, Ganeshkhind, Pune-411007, India}

\author[0000-0003-4864-2806]{Paolo Leto}
\author[0000-0002-6972-8388]{Grazia Umana}
\author[0000-0003-2876-3563]{Innocenza Busa}
\author[0000-0003-1856-6806]{Francesco Cavallaro}
\affiliation{INAF-Osservatorio Astrofisico di Catania, Via Santa Sofia 78, I95123 Catania, Italy}

\author[0000-0001-8704-1822]{Barnali Das}
\affiliation{Department of Physics and Astronomy, Bartol Research Institute, University of Delaware, 217 Sharp Lab, Newark, DE 19716, USA}

\author[0000-0002-0844-6563]{Poonam Chandra}
\affiliation{National Centre for Radio Astrophysics, Tata Institute of Fundamental Research, Ganeshkhind, Pune-411007, India}
\affiliation{National Radio Astronomy Observatory, 520 Edgemont Rd, Charlottesville, VA 22903, USA}

\author[0000-0001-5654-0266]{Miguel P\'erez-Torres}
\affiliation{Instituto de Astrof\'isica de Andaluc\'ia (IAA-CSIC),
Glorieta de la Astronom\'ia s/n, E-18008 Granada, Spain}
\affiliation{Facultad de Ciencias, Universidad de Zaragoza, Pedro Cerbuna 12, E-50009 Zaragoza, Spain} 

\author[0000-0002-1854-0131]{Gregg A. Wade}
\affiliation{Department of Physics and Space Science, Royal Military College of Canada, PO Box 17000, Station Forces, Kingston, ON K7K 7B4, Canada}

\author[0000-0002-7703-0692]{Cristobal Bordiu}
\author[0000-0002-7288-4613]{Carla S. Buemi}
\author[0000-0002-3429-2481]{Filomena Bufano}
\author[0000-0002-3137-473X]{Adriano Ingallinera}
\author[0000-0001-5126-1719]{Sara Loru}
\author[0000-0001-6368-8330]{Simone Riggi}
\affiliation{INAF-Osservatorio Astrofisico di Catania, Via Santa Sofia 78, I95123 Catania, Italy}

\begin{abstract}
In exoplanetary systems, the interaction between the central star and the planet can trigger Auroral Radio Emission (ARE), due to the Electron Cyclotron Maser mechanism. The high brightness temperature of this emission makes it visible at large distances, opening new opportunities to study exoplanets and to search for favourable conditions for the development of extra-terrestrial life, as magnetic fields act as a shield that protects life against external particles and influences the evolution of the planetary atmospheres.

In the last few years, we started an observational campaign to observe a sample of nearby M-type stars known to host exoplanets with the aim to detect ARE. We observed YZ\,Ceti with the upgraded Giant Metrewave Radio Telescope (uGMRT) in band 4 (550–-900\,MHz) nine times over a period of five months. We detected radio emission four times, two of which with high degree of circular polarization. With statistical considerations we exclude the possibility of flares due to stellar magnetic activity. 
Instead, when folding the detections to the orbital phase of the closest planet YZ\,Cet\,b, they are at positions where we would expect ARE due to star-planet interaction (SPI) in sub-Alfvenic regime. With a degree of confidence higher than $4.37\,\sigma$, YZ\,Cet is the first extrasolar systems with confirmed SPI at radio wavelengths. Modelling the ARE, we estimate a magnetic field for the star of about 2.4\,kG and we find that the planet must have a magnetosphere. The lower limit for the polar magnetic field of the planet is $0.4$\,G.
\end{abstract}


\section{Introduction} \label{sec:intro}

The presence of magnetospheres surrounding terrestrial planets is believed to play an important role in the evolution of the planetary atmospheres and in the development of life \citep{Griessmeier2005, Griessmeier2016, Owen2014, McIntyre2019, Green2021}. Magnetic fields act as a shield that prevents the arrival of ionized and potentially dangerous particles at the planetary surface \citep{Shields2016, Garcia2017}. This happened to the Earth, which has a magnetic field, and, among the planets in the habitable zone of the solar system, is the only one where life is known to have emerged.

On the other hand, intense solar flares and coronal mass ejections (CME) may compress the planet's magnetosphere causing the opening of the polar cups and providing a free way for energetic particles to precipitate in the atmosphere \citep{Airapetian2015, Airapetian2017}, producing fixation of molecules, as nitrogen and carbon dioxide and, possibly, ingredients for the development of life. This may have happened in atmosphere of the young Earth \citep{Airapetian2016}. In this context, both planetary magnetospheres and stellar activity, with increasing ionizing radiation (UV, X-rays) \citep{Lammer2012, Vidotto2022}, play important roles in creating a favourable environment for the development of life. In addition, the presence of a magnetic field in planets gives the opportunity to infer important characteristics of their interiors as an indicator of internal dynamo \citep{Lazio2019}.

The analysis of observations at radio wavelengths, which are sensitive to flares, associated energy releases and particles acceleration, is important to probe the interplanetary space in planetary systems other than the solar system.

So far, many planets have been found around red and ultracool dwarfs, which constitute the most common stars in our Galaxy and are the majority of nearby stars. They possess long-lived, suitable conditions for the development of life in their planetary systems. Earth-sized planets, some of them in the habitability zone, have been detected orbiting cool stars, as for example in the case of Trappist-1 \citep{Gillon2016, Gillon2017}, Proxima Cen \citep{Anglada-Escude2016} and Teegarden's Star \citep{Zechmeister2019}.

Aurorae are important manifestations of Star-Planet Interaction (SPI) in all the magnetized planets of the Solar System, detected as line emission in optical, UV and X-rays. These emissions are due to the precipitation of energetic charged particles of the solar wind in the planet's atmosphere around the polar magnetic caps. Moreover, the magnetic interaction with satellites in close orbit, as in the case of Jupiter and its Galilean moons, triggers particle acceleration that causes aurorae in the polar caps of the giant planet. At radio wavelengths, highly beamed, strongly polarized bursts are visible. They appear to originate from an annular region above the magnetic poles, associated to auroras in the atmosphere of Jupiter \citep[e.g.][]{Zarka1998}. This is interpreted in terms of Electron Cyclotron Maser Emission (ECME) that originates in the magnetospheric auroral cavities, and is called Auroral Radio Emission (ARE).

\section{Auroral Radio Emission}

The ECME is a coherent emission mechanism due to the gyro-resonance of an asymmetric population of electrons in velocity space. This can occur when electrons converging toward a central body, following the magnetic flux tubes, are reflected back by magnetic mirroring. Since electrons with small pitch angles penetrate deeper, they precipitate in the atmosphere of the central body, causing ultraviolet and optical auroras. This leads to a loss-cone anisotropy in the reflected electronic population, i.e. an inversion of population in velocity space, giving rise to maser emission. This amplifies the extraordinary magneto-ionic mode, producing almost 100\% circularly polarized radiation at frequencies close to the first few harmonics of the local gyro-frequency ($\nu_{\mathrm B}=2.8 B$ MHz, with $B$ in G). Locally, the amplified radiation is beamed in a thin {\it hollow cone}, whose axis in tangent to the local magnetic field line ({\it hollow cone model}) \citep{MelroseDulk1982}. 

ARE is also observed in single stars, as hot magnetic chemically peculiar stars (mCP) \citep[e.g.][]{Trigilio2000, Das2022, Leto2020}, and in many very low mass stars and Ultra Cool Dwarfs (UCDs), with spectral type ranging from M8 to T6.5 \citep[e.g.][]{Berger2009, Hallinan2007, Route2012, Lynch2015}.  Notwithstanding that they are located in very different regions of the Hertzsprung-Russell (HR) diagram, these stars have a common characteristic: a strong magnetic field, dominated by the dipole component, tilted with respect to the star’s rotational axis. In mCP stars, where the magnetic topology is known, we observe two pulses at two rotational phases, close to the moments where the axis of the dipole lies in the plane of the sky. As the star rotates, the ECME produces a lighthouse effect, similar to pulsars. The same behaviour is observed in a few UCDs \citep{Hallinan2007}. In Solar system planets, the location of the origin of ECME, as in the case of the auroral kilometric radiation (AKR) of the Earth \citep{Mutel2008}, is the same as that derived from observations of stars, i.e. at a height of about $0.1-2$ stellar radii above the poles, tangent to annular rings of constant B. This is in agreement with the {\it tangent plane beaming model} \citep{Trigilio2011}. This pattern of emission can occur when it originates in all points of the annular ring, each of them with a hollow cone pattern,  and the overall emission is the sum of the emission from each ring; in the tangential direction the radiation is intensified. 
On the contrary, the {\it hollow cone model} seems more adequate when the maser acts only in a small portion of the annular ring, corresponding to the flux tube connecting the planet, and the emission pattern is the natural hollow-cone. This pattern explains the ARE in most Solar system planets and is invoked to explain the radio emission arising from exo-planets.
However, for both models, ARE is foreseen to appear in symmetric orbital position of the planet with respect to the line of sight.

There are two kinds of ARE due to the interaction between our Sun and planets, which are believed to also act in exoplanetary systems. The first is due to the ram pressure of the wind of the star on the magnetosphere of the planet. In this case the frequency of the ECME is proportional to the magnetic field strength of the planet ($B_\mathrm{planet}$) for which any detection of ARE provides a direct measurement. However, since $B_\mathrm{planet}$ is expected to be of the order of a few gauss, the frequency of the maser is expected to fall at the edge, or below, the ionospheric boundary of the radio window.  In fact, the search for this emission gives basically negative results \citep[e.g.][]{Bastian2000, Ryabov2004, Hallinan2007, Lecavelier2013, Sirothia2014}.

The second kind is due to the interaction of the orbiting planet with the magnetosphere of the parent star. This case is analogous to the system of Jupiter and its moons. At the present, there are some possible detections of this kind of ARE.  The observed features in the time-frequency domain of the stellar ARE from the M8.5-type star TVLM\,513-46546 \citep{Hallinan2007, Lynch2015} were explained as a signature of an external body orbiting around this UCD. This possibility is supported by a model developed by \cite{Leto2017}. \cite{Vedantham2020} claimed the detection of ARE from GJ\,1151, an M4.5V star at 8.04\,pc, by comparing two observations made during the LOFAR Two-Metre Sky Survey (LoTSS, \citealt{Tasse2021}). They detected Stokes V on one epoch, suggesting a possible SPI between the star and a hypothetical planet in close orbit. Indeed, \cite{Mahadevan2021} report the possibility of a planet of 2-day orbit, but \cite{Perger2021} ruled out this hypothesis with accurate radial velocity measurements. Similarly, \cite{Davis2021} report possible ARE in the dMe6 star WX\,UMa by comparing three observations of the LoTSS survey. However, none of these observations demonstrate that this ECME is due to SPI, since no planets have been found around these stars. The only successful way to associate ECME with SPI is to observe stars with confirmed planets for which orbital parameters are known, looking for a correlation of any detected ECME with the orbital phase or with periodicity in the radio emission different form the rotation rate of the star. This has been attempted by \cite{Trigilio2018} who observed $\alpha$\,Cen\,B with the aim to detect ARE from  $\alpha$\,Cen\,Bb, \citep{Dumusque2012}. However, no detection has been reported; moreover, in this case the presence of a planet was ruled out \citep{Rajpaul2016}. 
The most evident case of ARE from SPI is that of the Proxima\,Cen - Proxima\,Cen\,b system, which was observed by
\cite{Perez-Torres2021} in the 1-3 GHz band with the Australia Telescope Compact Array (ATCA) in 2017 for 17 consecutive days (spanning $\sim$1.6 orbital periods). They detected circularly polarized radio emission at 1.6 GHz at most epochs, a frequency consistent with the expected electron-cyclotron frequency for the known star’s magnetic field intensity of $\sim$600 gauss \citep{Reiners2008}. Based on the 1.6 GHz ATCA light curve behavior, which showed an strongly circularly polarized emission pattern that correlated with the orbital period of the planet Proxima b, \cite{Perez-Torres2021} found evidence for auroral radio emission arising from the interaction between the planet Proxima b and its host star Proxima.

With the aim to search for additional robust detections of ARE due to SPI, we started an observational campaign with several radio interferometers. The targets are nearby exoplanetary systems around late-type stars with planets in close orbit. In this Letter, we report the results of one of these campaigns, carried out with the uGMRT, which resulted in the detection of highly-polarized radio emission from YZ\,Ceti, which is consistent with ARE due to SPI between the planet YZ\,Ceti b and its host star.

\section{YZ Ceti}
\label{yzcet}

YZ\,Cet (GJ 54.1, 2MASS J01123052-1659570) is an M4.5V type star with a mass $M_\ast=0.14\,M_\odot$ and a radius $R_\ast=0.157\,R_\odot$ \citep{Stock2020}, at a distance of 3.71 pc \citep{Gaia2018}, hosting an ultra-compact planetary system. At the present time, three Earth-mass planets have been discovered with the radial velocity (RV) method \citep{Astudillo-Defru2017}, namely YZ\,Cet b, c, d with orbital periods $P_\mathrm{orb}=2.02,3.06,4.66$ days and semi-major axes $r_\mathrm{orb}=0.016, 0.022, 0.028$ au, respectively \citep{Stock2020}, corresponding to $21.9, 30.1, 38.3\, R_\ast$. No planetary transits have been observed for the YZ\,Cet system. For this reason the radii of the planets are not measured, but there is an estimate of $R_\mathrm{b}=0.93$, $R_\mathrm{c}=1.05$ and $R_\mathrm{d}=1.04\,R_\oplus$ from a semi-empirical mass-radius relationship \citep{Stock2020}.
~\\
YZ\,Cet is a mid-M type star classified as an eruptive variable. Stars of this spectral type tend to have strong, kG, axisymmetric dipolar field topologies \citep[and references therein]{Kochukhov2017}. YZ\,Cet is a slow rotator, with a period $P_\mathrm{rot}=68$ days \citep{Stock2020} and an age of 3.8\,Gyr \citep{Engle2017} and from the activity indicator, based of the H\&K CaII UV lines, $\log R^{'}_\mathrm{HK}=-4.87$ we deduce that it has a low activity level \citep{Henry1996}.

The coronal X-ray luminosity determined from two ROSAT measurements is Lx$\approx 10^{27.1}\,\mathrm{erg\,s^{-1}}$ similar to the solar value (Lx$_\odot \approx 10^{26.8}-10^{27.3}\,\mathrm{erg\,s^{-1}}$, \citealt{Judge2003}).
From the Guedel Benz relation \citep{Guedel1993}, coupling X-ray and spectral radio luminosities in stars (Lx$\approx$Lr,$_\nu\times10^{15.5}$), we can estimate the basal radio luminosity of YZ\,Cet (Lr$\approx10^{11.6}\,\mathrm{erg\,s^{-1}Hz^{-1}}$) and, assuming a distance of 3.71\,pc, a basal radio flux density  of $S_\nu\approx 25\,\mu$Jy.
YZ\,Cet was observed several times at radio wavelengths, from 843 to 4880\,MHz \citep{Wendker1995, McLean2012}, but never detected.
\\
\cite{Vidotto2019} assert that YZ\,Cet\,b could give detectable ARE, due to interaction of the stellar wind with the planet’s magnetosphere, but at MHz frequencies.
Very recently\footnote{after this paper was initially submitted}, \cite{Pineda2023}, observed YZ\,Cet with the VLA at 2-4\,GHz in five days, from Nov 2019 to Feb 2020. They detected two coherent bursts with an high degree of circular polarization, modeling their results as due to ARE from SPI, but not excluding the possibility of flares due to stellar magnetic activity.

\begin{table}[t]
\begin{center}
\caption{Log of the uGMRT Observations 
\label{tab:observations}}
\begin{tabular}{cclc}
\hline
\hline
Day        & Date & Start - Stop       & Duration \\
number     &      & ~~~~~~UT           & min  \\
\hline
  1 & 01-May-2022 & 04:33 - 05:36      & 63 \\
  2 & 16-May-2022 & 04:46 - 05:24      & 38 \\
  3 & 03-Jun-2022 & 02:53 - 03:27      & 34 \\
  4 & 15-Jun-2022 & 00:38 - 01:47      & 69 \\
  5 & 01-Jul-2022 & 22:55 - 00:03 (+1d)& 68 \\
  6 & 15-Jul-2022 & 23:34 - 00:27 (+1d)& 53 \\
  7 & 02-Aug-2022 & 00:23 - 00:40      & 17 \\
  8 & 14-Aug-2022 & 22:50 - 23:58      & 68 \\
  9 & 03-Sep-2022 & 20:15 - 21:20      & 65 \\
\hline
\end{tabular}
\end{center}
\end{table}

\begin{table}[t]
\begin{center}
\caption{Results \label{tab:results} 
}
\begin{tabular}{cccccc}
\hline
\hline
Day        & Phase  & Stokes I  & rms      & Stokes V & rms \\
number     &        & mJy       & mJy      & mJy      & mJy \\
\hline
   1       & 0.610  & --        &  0.050   &          &       \\  
   2       & 0.032  & --        &  0.047   &          &       \\ 
   3       & 0.881  & 0.420     &  0.052   & --       & 0.023 \\ 
   4       & 0.797  & 0.290     &  0.053   & --       & 0.026 \\ 
   5       & 0.176  & 0.623     &  0.047   & +0.580   & 0.027 \\ 
   6       & 0.114  & 1.070     &  0.060   & +0.800   & 0.050 \\ 
   7       & 0.537  & --        &  0.070   &          &       \\ 
   8       & 0.947  & --        &  0.045   &          &       \\ 
   9       & 0.790  & --        &  0.100   &          &       \\ 
\hline
\end{tabular}
\end{center}
\end{table}

\begin{figure}[t]
\includegraphics[width=8.5cm]{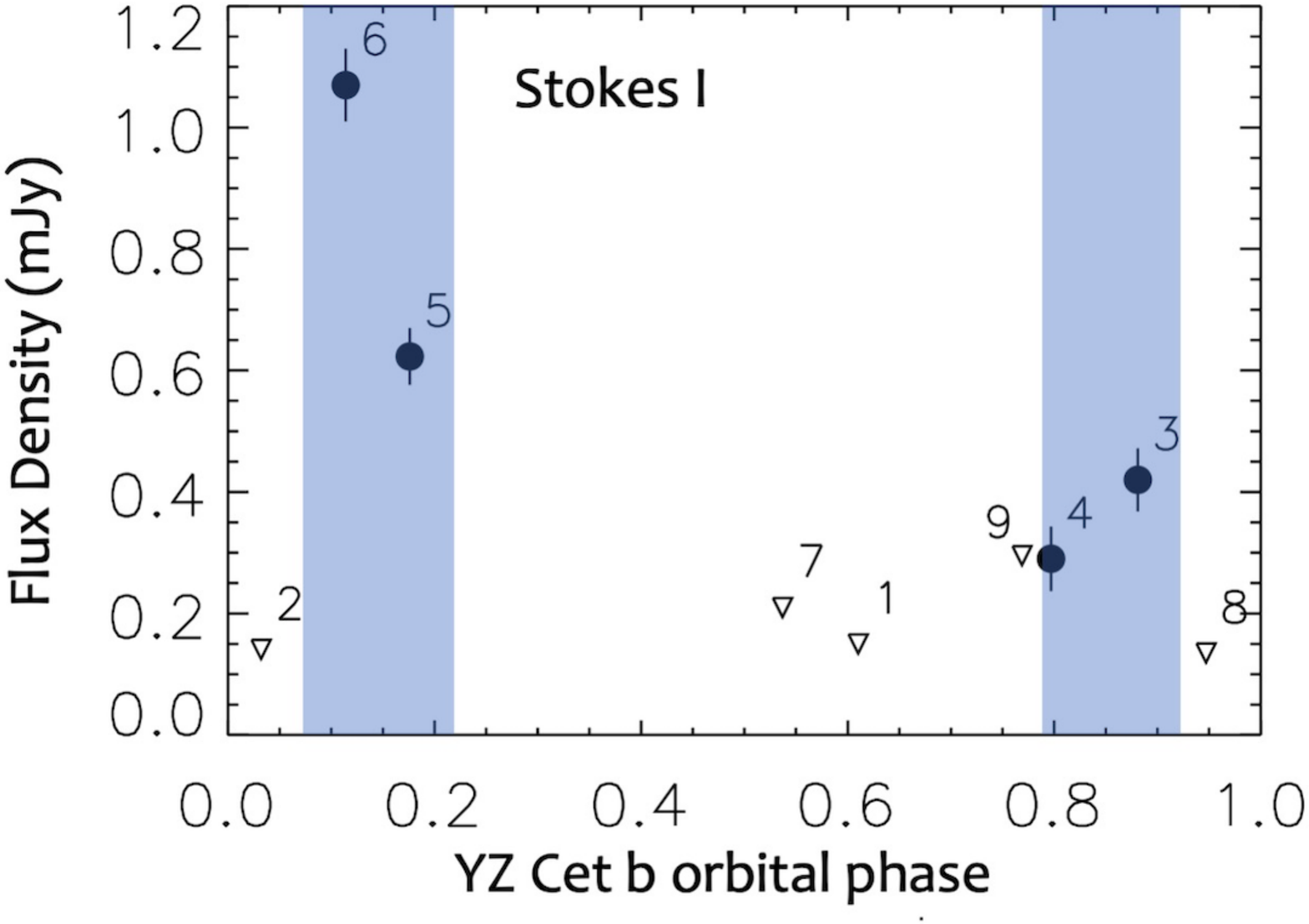}
\caption{Radio light curve at band 4 towards the system YZ\,Cet as a function of the orbital phase of planet b. Detectable flux density is visible at two ranges of phases (blue areas). Upper limits are shown by open downward triangles. 
\label{fig:lightcurve}}
\end{figure}
\begin{figure}[ht]
\includegraphics[width=8.5cm]{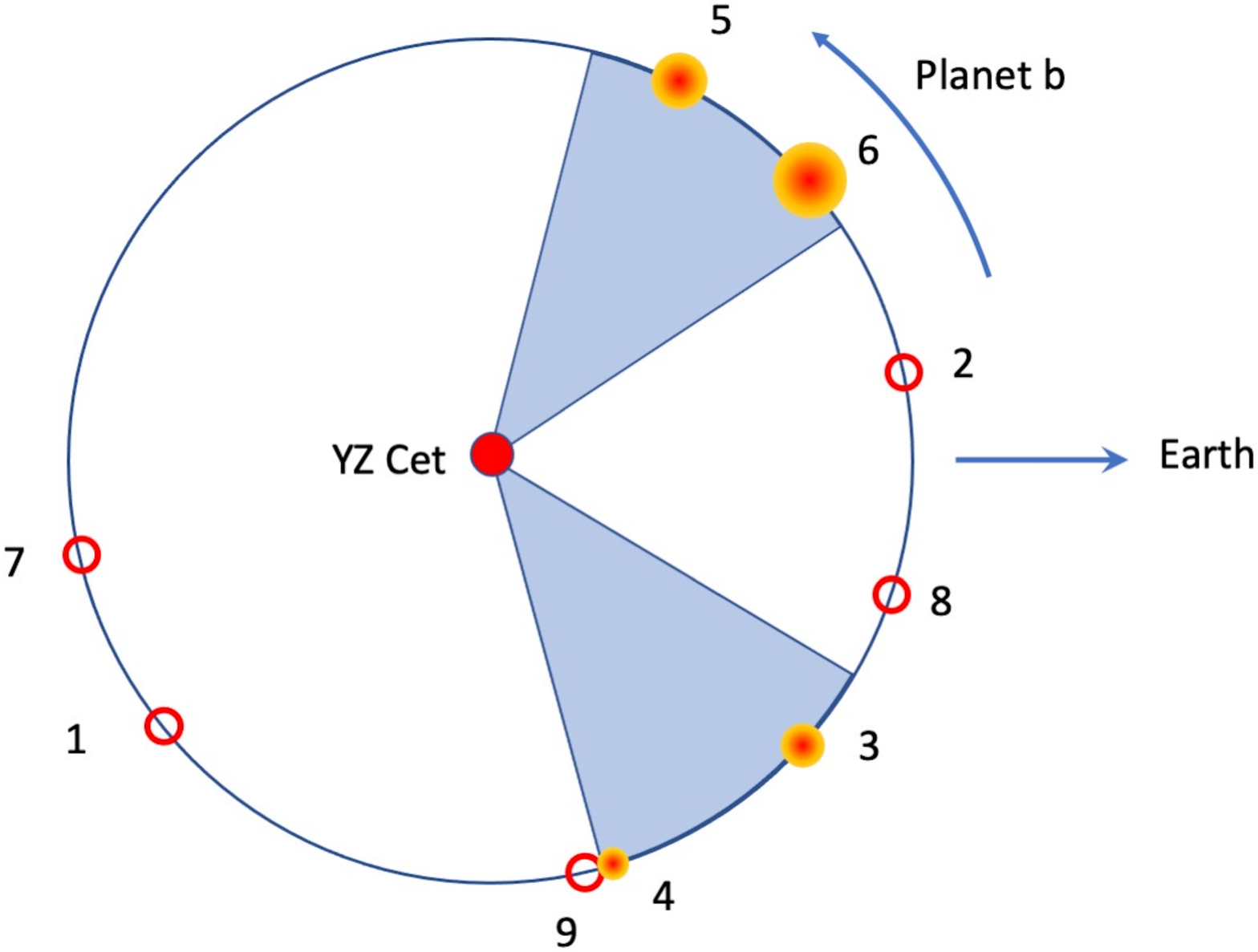}
\caption{Schematic view of the detected radio emission as a function of the position of planet b in the orbit. YZ\,Cet is in the center, the direction of the Earth is on the right. Detection is marked by red-orange filled circles, whose size is proportional to the flux density. Non-detections are marked by red empty circles.
\label{fig:orbit}}
\end{figure}

\section{Observations and data reduction}
We observed YZ\,Cet with the upgraded Giant Metrewave Radio Telescope (uGMRT) \citep{Gupta2017} at band 4 (550–900\,MHz), using 3C48 (J0137+331) as amplitude and bandpass calibrator and J0116-208, which is about 4$^{\circ}$ from the target, as the phase calibrator. YZ\,Cet was observed  9 times from May to September 2022, together with the calibrators. Logs of the observations are reported in Table\,\ref{tab:observations}. There is no particular periodicity in the days of observations, which are essentially randomly distributed.\\
Data have been flagged, calibrated and mapped using the Common Astronomy Software Applications (\textsc{casa}) \citep{McMullin2007}. 
We used a Python-based pipeline developed for continuum imaging in CASA for point source observations with uGMRT (Biswas et al., in prep.). The scripts use the CASA task `\textsc{flagdata}' and automatic flagging algorithm `\textsc{tfcrop}' to remove radio frequency interference (RFI). The central baselines were treated with extra precaution to improve data quality. The calibration process was done in several iterations with conservative flagging to reduce the amount of flagged data. In the first step, the calibration solutions were applied to the flux calibrator only, and the calibrated data were flagged using another automatic RFI excluding algorithm `\textsc{rflag}'. Although initially, a wider band of data was taken for analysis according to the data quality, after the first iteration, only a fixed final bandwidth of 265 MHz was used. In consecutive iterations, new calibration solutions were applied to the phase calibrators and the target, respectively. At each step, minimum signal-to-noise for calibration steps and flagging parameters were changed. This method improved the calibration, while keeping the flagging percentage as low as possible. Averaging in the frequency space on the final data were performed to obtain a final spectral resolution of 0.78 MHz. All the imaging were done using CASA task `\textsc{tclean}', with deconvolver `\textsc{mtmfs}' (Multiscale Multi-frequency with W-projection, \citealt{Rau2011}). Several rounds of phase self-calibration steps were performed to improve the imaging results using the `\textsc{gaincal}' \& `\textsc{tclean}' tasks. To remove the strong imaging artefacts created by bright sources near the phase centre, some of the nearby bright sources were removed from the visibility plane. This was done by subtracting the model visibilities of those corresponding bright nearby sources using task `\textsc{uvsub}'. Finally, several rounds of phase-only and two rounds of amplitude and phase (A \& P)-type self-calibration were performed to get the final radio image.\\
Analysis of the maps has been carried out using the task \textsc{imfit} to measure the integrated flux density of the source assuming a two-dimensional Gaussian and \textsc{imstat} for the evaluation of the RMS of the maps near the target.
Stokes I data were analysed for all the days of observations, whereas Stokes V data were analysed only in the case of detection. Results of the analysis are provided in Table\,\ref{tab:results}.

\section{Results}

YZ\,Cet has been detected four days out of nine, namely days 3, 4, 5 and 6 (Table\, \ref{tab:observations}). 
Stokes I is between 290 and 1070 $\mu$Jy, more than 5$\,\sigma$ for the lowest emission that occurs on day 4. Stokes V is reported only on days 5 and 6, with positive values on both days\footnote{The uGMRT sign convention in band 4 for defining right and left circular polarization is opposite to the IAU/IEEE convention (Das et al. 2020). We have taken this fact into account in our post-processing of the uGMRT data so that the convention used in this work is the same as the IAU/IEEE convention for circular polarization}, and with a very high percentage of circular polarization, 93\% and 75\% respectively. The highly circularly polarized radio emission in days 5 and 6 is consistent with the ECME.

\begin{figure}[ht!]
\includegraphics[width=8cm]{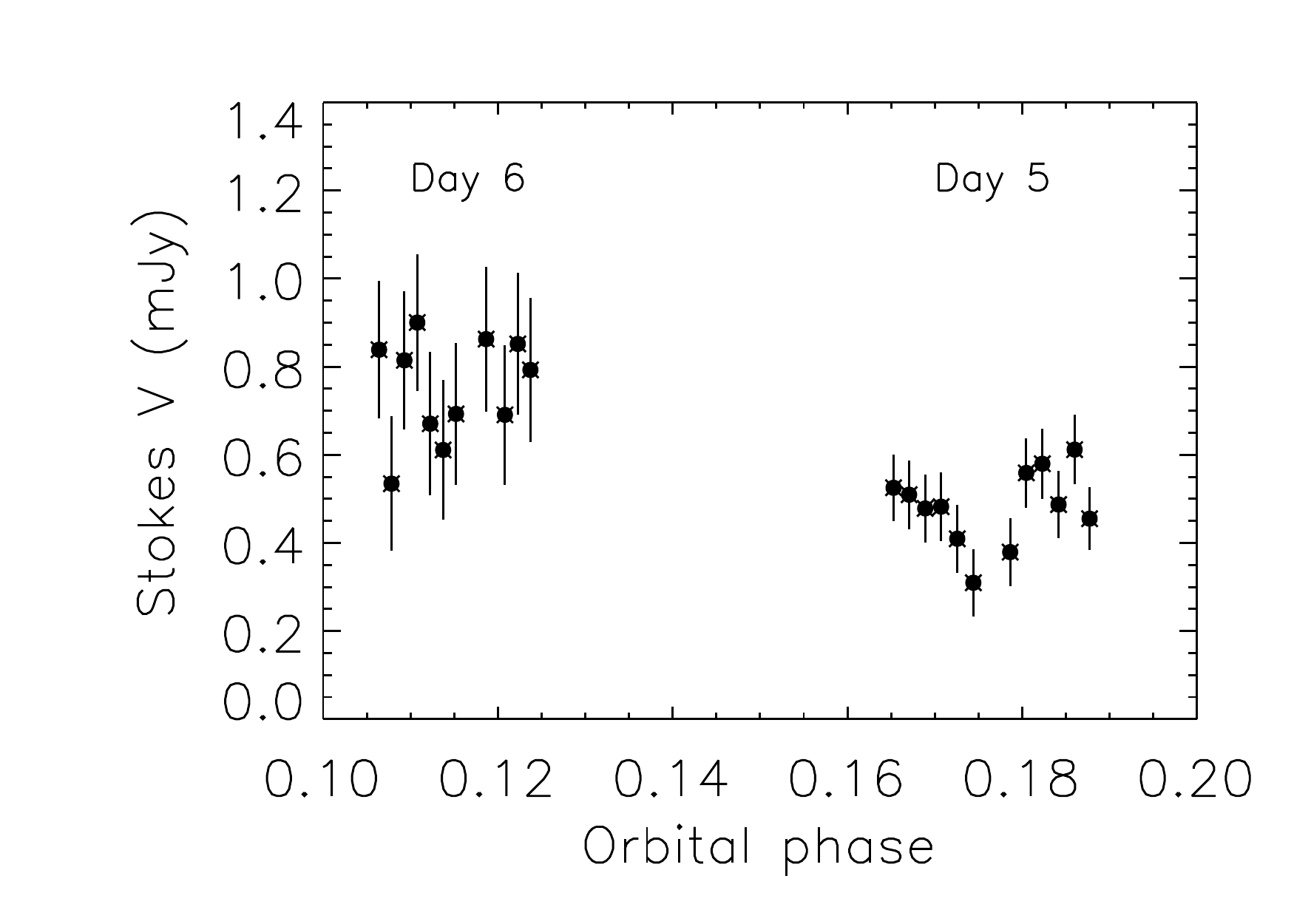}
\caption{Light-curve of the circularly polarized component of ARE in the two days of detection, folded to the orbital period of planet b.}
\label{fig:curvamedia}
\end{figure}

\begin{figure}[ht!]
\includegraphics[width=8.cm]{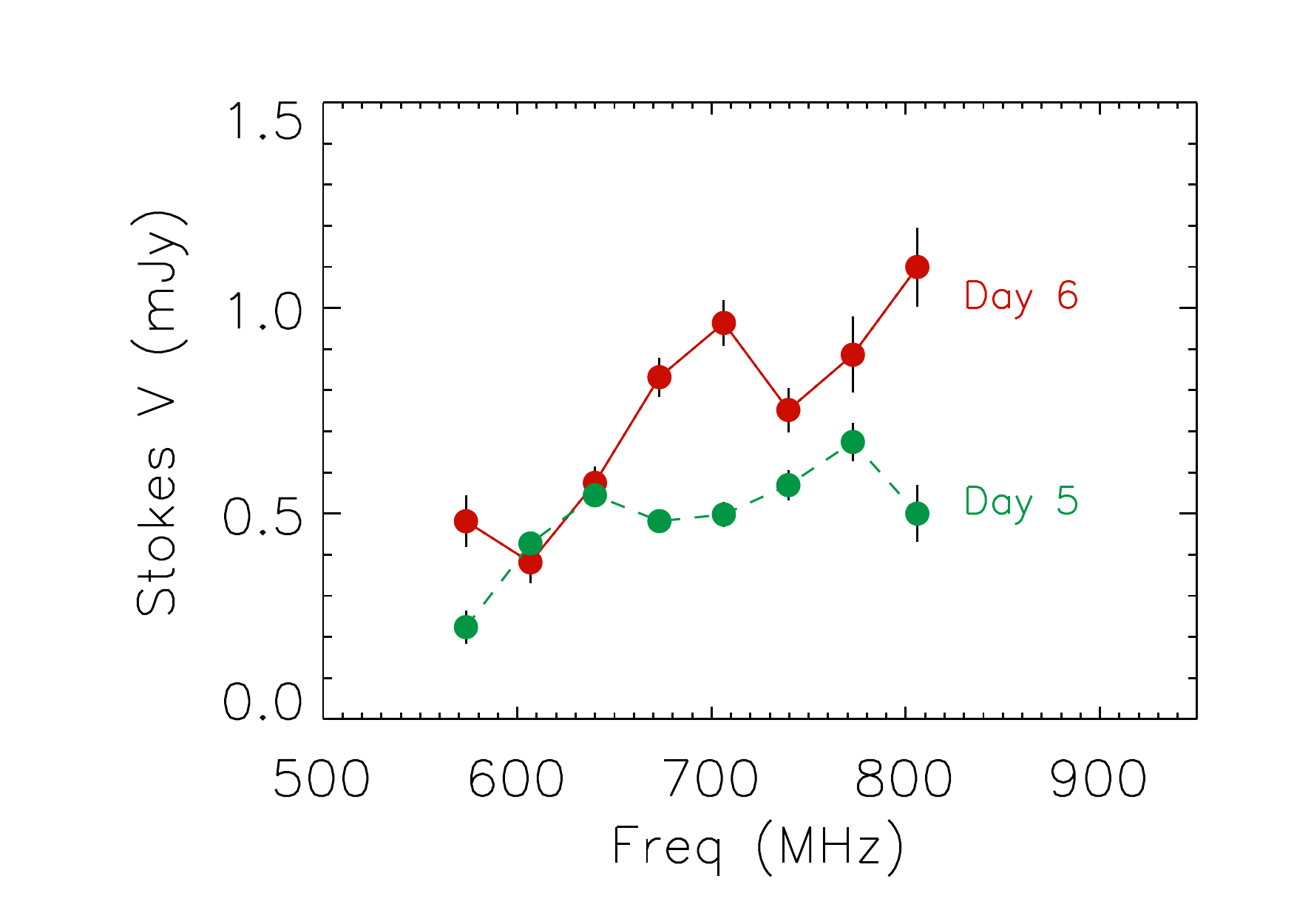}
\caption{Spectra of the circularly polarized component of ARE in the two days of detection. The spectrum is increasing towards high frequency, indication that the high frequency cutoff is beyond the limits of the figure.}
\label{fig:spettromedio}
\end{figure}

In order to investigate the presence of SPI, the Stokes I flux densities have been folded with the orbital periods of the three known planets. We used the ephemeris provided by \citet{Stock2020} that, for planet b, are: $HJD=2452996.25$ and $P_\mathrm{orb}=2.02087$\,days.
While for planet c and d, detections and non detections are randomly mixed around the orbits, for planet b the phases corresponding to detections appear in two groups, approximately between phases 0.07--0.2 and 0.78--0.9. The two intervals are marked as light blue areas in Fig.\,\ref{fig:lightcurve}, and the corresponding orbital positions are shown in Fig. \ref{fig:orbit}.

A deeper analysis has been carried out in Stokes V for the two days of detection. We computed a spectrogram of the emission by performing the Discrete Fourier Transform (DFT) of the complex visibilities at the position of the star as a function of time and frequency channels. This analysis was carried out only for Stokes V since there are no other sources in the field, while in Stokes I this analysis suffers from the presence of sidelobes of other sources at the position of the target.
The dynamical spectra do not show any notable structures. We then obtained light curves by averaging first over the whole bandwidth and then with a time resolution of 4 minutes. These are shown in Fig.\,\ref{fig:curvamedia}, where time is converted into orbital phase. 
During Day 6 the temporal behaviour of Stokes V is a little noisier with respect to that of day 5, and does not show any particular trend. During day 5 it is possible to appreciate a decrease of emission at a middle of the observation, demonstrating that the emission is likely not  constant even on short timescales ($\lessapprox1$ hour). \\
We also obtained in-band spectra for days 5 and 6 by averaging first in the whole time-range and then with a resolution of 33\,MHz. In-band spectra are shown in Fig.\,\ref{fig:spettromedio}. During day 5 the flux density increases in the first part of the band ($550-650$\,MHz) then it is almost flat. During day 6 it increases on average, indicating that the spectrum probably extends to higher frequencies, with a possible cutoff at more than 1\,GHz. For both days, the steep increase of the flux density seems to point to a minimum frequency of about 500\,MHz, which could indicate a low limit of the ECME. 

\begin{figure}[t]
\includegraphics[width=8.5cm]{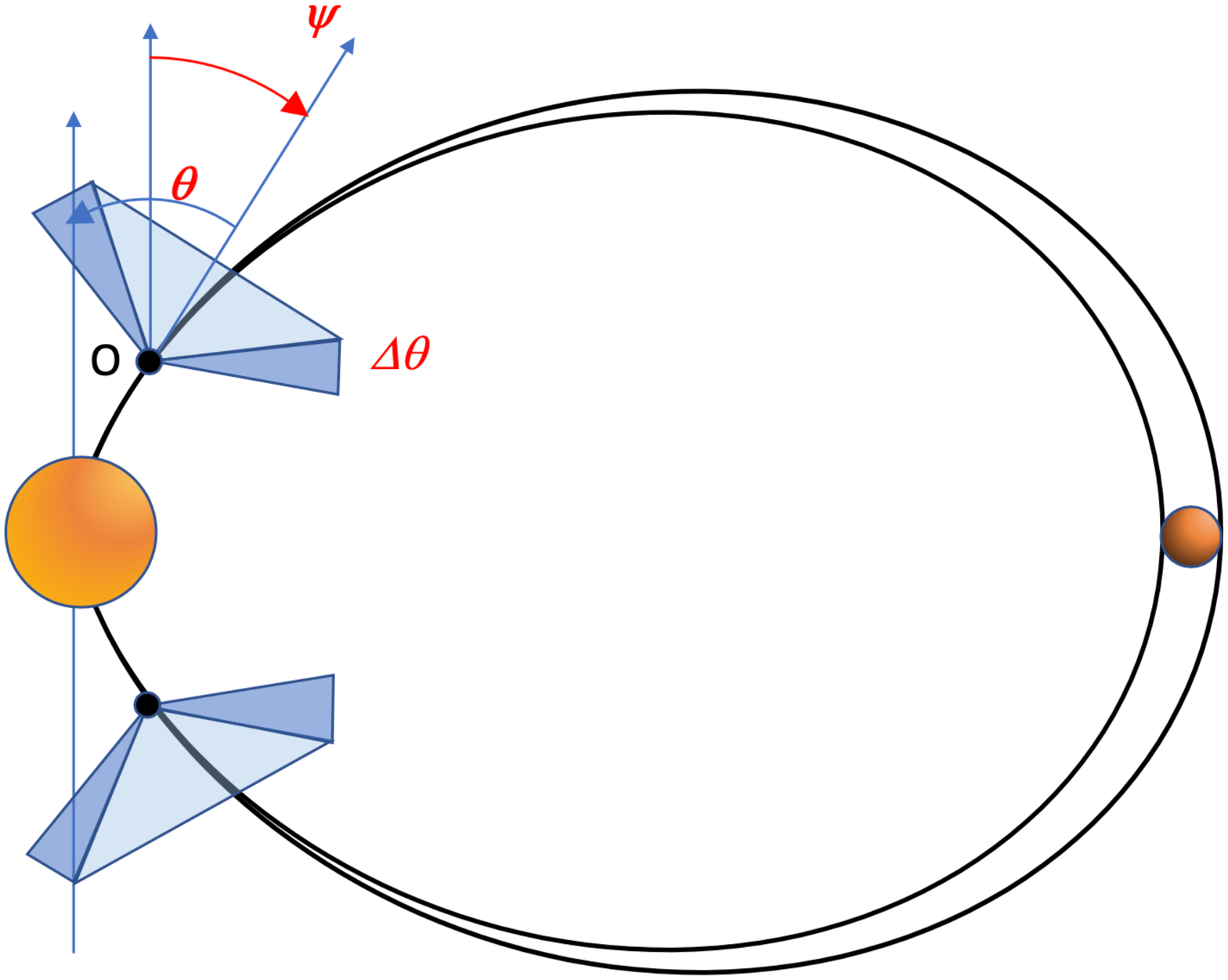}
\caption{Schematic view of the magnetic connection between star and planet. The stellar dipole is assumed to be perpendicular to the orbital plane. The axes of the ECME cones are tangent to the dipole line (angle $\psi$ to the dipole axis). The aperture of the cone is $\theta$ with thickness $\Delta\theta$, centered in O. 
\label{fig:dipolo}}
\end{figure}

\section{Discussion}
\subsection{Is the emission really due to SPI?}
The orbital phases of the four detections define two sectors (blue areas in Fig.\ref{fig:lightcurve} and \ref{fig:orbit}) that are symmetric with respect to the line of sight. The two sectors cover $\pm(30^{\circ}$ to $80^{\circ}$), with a total of $100^{\circ}$ over $360^{\circ}$.
This symmetry strongly suggests that we are detecting ARE due to SPI in the magnetosphere of the star due to sub-Alfv\'enic interaction with planet b, that can be explained in the framework of the hollow cone model. 
However, other emission mechanisms observed in M type stars could be responsible for the observed emission.
%
The radio emission from active M stars is highly variable and is characterized by the presence of two kinds of flares superimposed to a quiescent radio emission. Incoherent flares are transient increases of radio flux, usually weakly circularly polarized, with timescales of order of hours,  modelled within the framework of gyrosynchrotron emission from mildly relativistic electrons \citep{Osten2005}. The other kind of flares  are coherent radio bursts, characterized by high level of circular polarization \citep{Villadsen2019} and timescales from seconds to hours. These characteristics are interpreted as ECME \citep{Lynch2015,Zic2019}.
At these days, the rate of coherent bursts in M type stars is still unknown. Only for the most active, fast rotating M type dwarfs, as AD\,Leo, UV\,Cet,  EQ\,Peg, EV\,Lac and YZ\,Cmi, \cite{Villadsen2019} have been able to estimate a rate of 20\% to catch a coherent burst at the same frequency of our observations. On the other hand, from a blind sky survey at low frequency ($\leq 200$\,MHz) \cite{Callingham2021} found a low rate of detection of coherent emission in M type stars, about $0.5\%$. This emission seems to be uncorrelated with the activity indicators while, at GHz frequency, there is a correlation with the Rossby number \citep{McLean2012}, i.e. with the magnetic activity. On the other hand, incoherent flares due to gyrosynchrotron emission have no suitable statistics for M type stars.

In any case, whatever the probability $p$ of flares or coherent bursts, 
the overall probability to get 4 detections inside the two sectors and other 5 non detections outside them is given by $p^4(1-p)^5$ which has a maximum around $2\times10^{-3}$. This is a very low probability. This occur when $p=0.44$, which is a very high flare rate, not suitable for the activity of YZ\,Cet. 
The two coherent burst reported by \cite{Pineda2023} can be used as a test. Phasing their data with the ephemeris we used, we get that their two detections occur at phase 0.13 and 0.09 (Epochs 2 and 5), which fall inside our sector; the non detections at phase 0.63, 0.62 and 0.76 (Epochs 1,3 and 4), which are outside our sectors. Considering all the data, 6 inside the sectors, and 8 outside, we get that the probability that flares or coherent bursts fall in this configuration is given by $P_\mathrm{tot}=p^6(1-p)^8$, which has a maximum of $7\times 10^{-5}$, corresponding to $4.37\,\sigma$, for $p=0.43$.

We can conclude that the observed emission is ARE from SPI, with a degree of confidence of $(1-\mathrm{max}(P_\mathrm{tot}))$, i.e. $99.992\%$.

\begin{figure}[t]
\includegraphics[width=8.5cm]{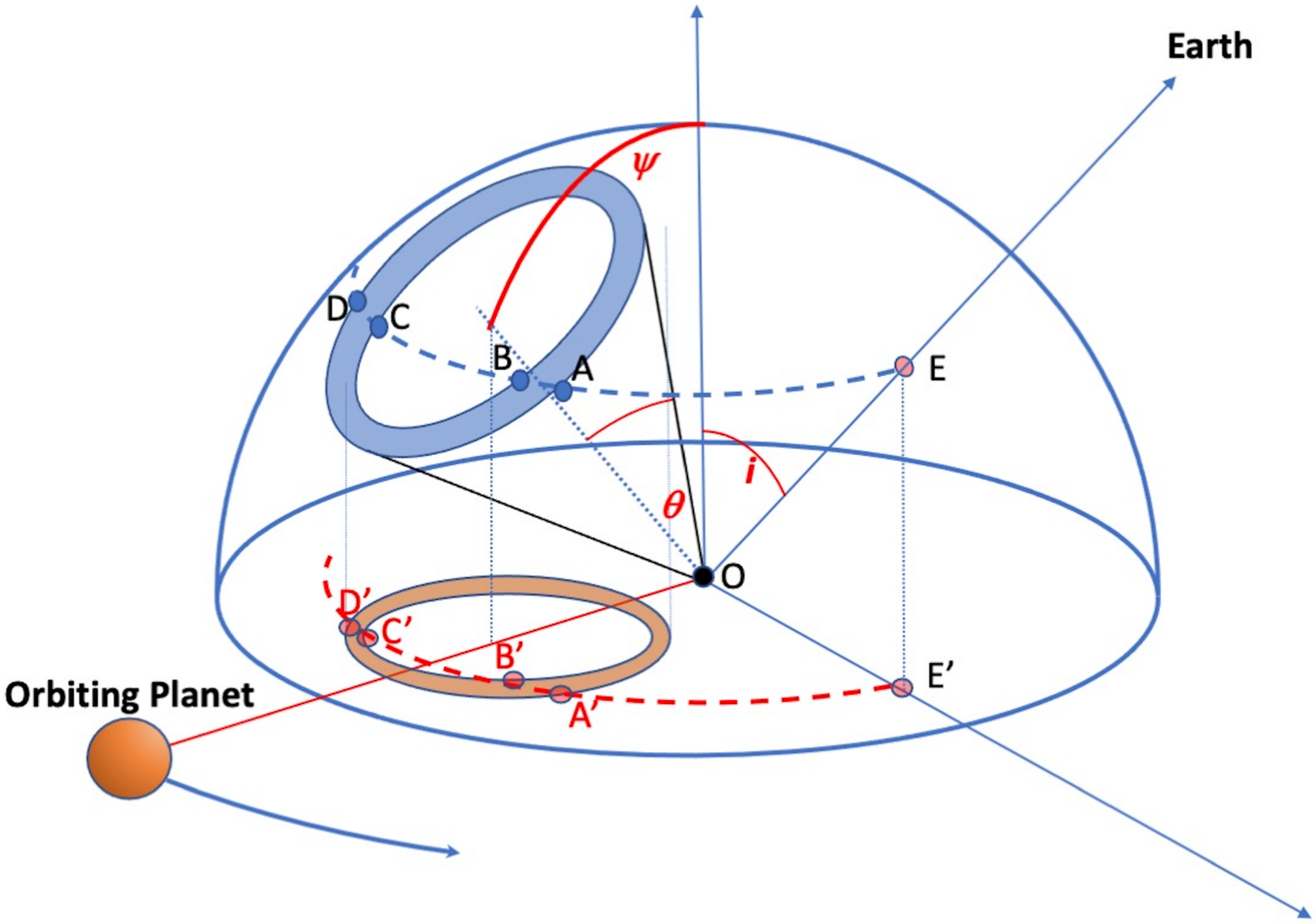}
\caption{Schematic view of the emission pattern of the ARE. Directions are projected in a unitary spherical surface. The ECME pattern is a hollow cone, represented by the blue ring on the sphere. The emission originates in the dipolar field line connecting the planet and the star in the point O and it rotates following the planet. The radiation is direct toward the Earth when the line of sight (point E) intercepts the cone (blue circular corona) between points A-B and C-D. This occurs when the projection E$'$ of E in the plane of the orbit  intercepts the projection of the blue ring (the orange ellipse) between points A$'$-B$'$ and C$'$-D$'$. 
\label{fig:cono}}
\end{figure}
\begin{figure}[t]
\includegraphics[width=7.cm]{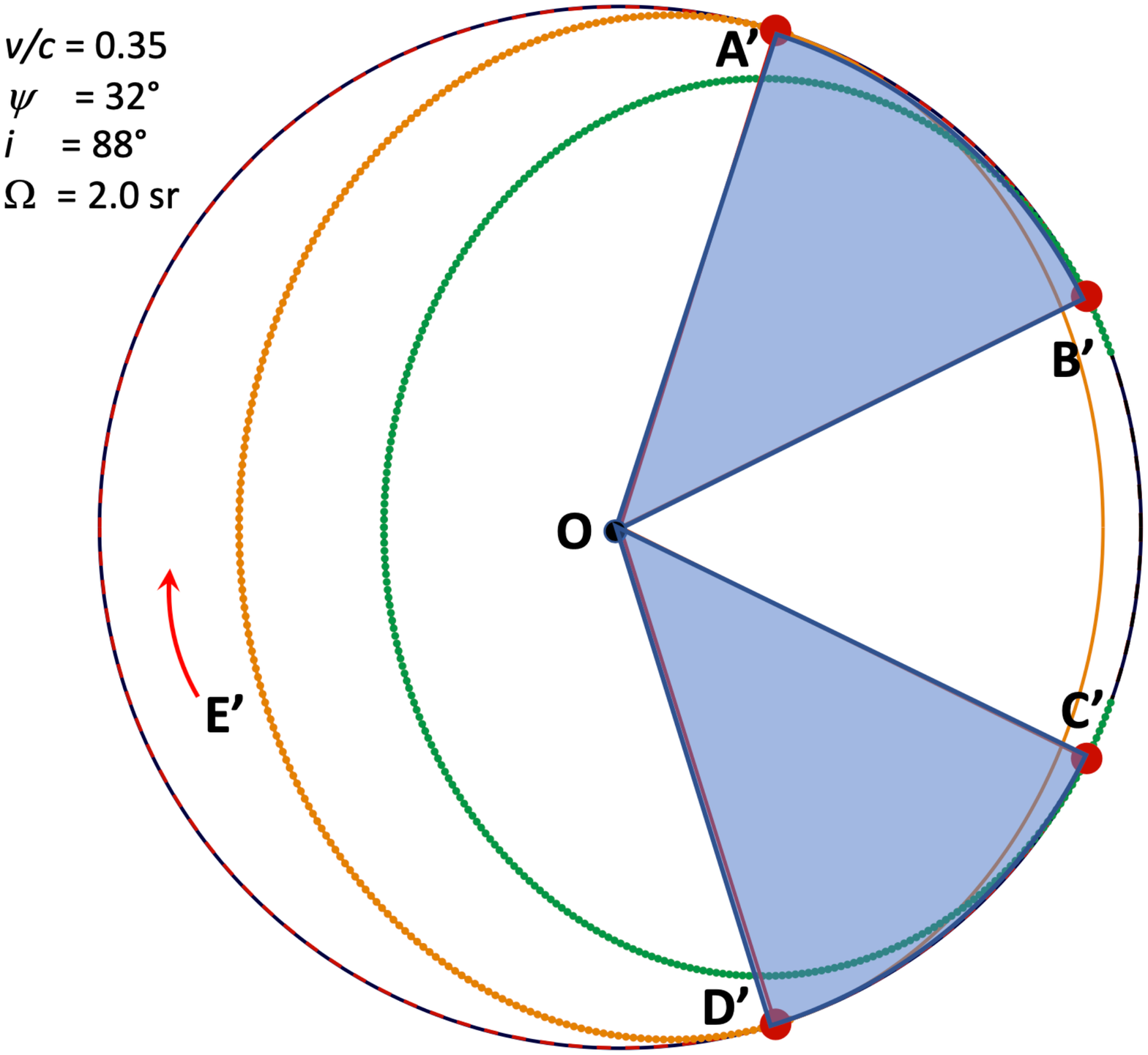}
\caption{Projection of one possible configuration in the plane of the orbit (horizontal plane of Fig.\,\ref{fig:cono}). The projection of the hollow cone is the area between the orange and the green ellipses. The projection of the line of sight (point E$'$) rotates and intercepts the two ellipses between points A$'$-B$'$ and C$'$-D$'$. Here $i=88^\circ$ and E$'$ describes a circle of radius $\approx 1$. ARE is visible from Earth when E$'$ in between A$'$ and B$'$ or C$'$ and D$'$. O is the origin of the emission.
\label{fig:cone-projection}}
\end{figure}
 
\subsection{Sub-Alfv\'enic regime}

The perturbation caused by the planet crossing the stellar magnetosphere can propagate towards the star if the relative velocity $v_\mathrm{rel}$ (see sect.\,\ref{planetaryfield}) of the planet with respect to the magnetosphere is less than the Alfv\'en velocity, given by $v_\mathrm{Alf}=B/\sqrt{4\pi\rho_\mathrm{w}}$, where $\rho_\mathrm{w}$ is the density of the wind \citep[e.g.][]{Lanza2009}. 
The value of $v_\mathrm{Alf}$ depends on the configuration of the magnetosphere, as it influences the density of the wind and therefore the ram pressure. Defining $\eta(r)=\frac{B^2/8\pi}{1/2\rho v_\mathrm{w}^2}$ the ratio between magnetic to wind energy densities, the Alfv\'en radius $R_\mathrm{Alf}$ is where $\eta(r)=1$. \cite{ud-Doula2002} define a "wind magnetic confinement parameter" $\eta\ast=B_\mathrm{P}^2R_\ast^2/4\dot M v_\mathrm{w}$, which is $\eta(R_\ast)$ at the stellar surface. If $\eta\ast \gg 1$, $R_\mathrm{Alf}\gg R\ast$, relatively far from the star. We can define ''inner magnetosphere'' the region where $R<R_\mathrm{Alf}$; here the magnetic field lines are closed \citep[as for mCP stars, see][]{Trigilio2004}. In the equatorial plane, assumed coincident with the orbital plane, $\eta(r)$ is the local ratio $(v_\mathrm{Alf}/v_\mathrm{w})^2=M_\mathrm{A}^{-2}$, with $M_\mathrm{A}$ the Alfv\'enic Mach number \citep{ud-Doula2002}. Inside the inner magnetosphere, $v_\mathrm{w}\ll v_\mathrm{Alf}$.\\
For mid-M type star with moderate or low activity, as YZ\,Cet, \citet{Wood2021} find that $\dot M\le 0.2\,\dot M_\odot$. Adopting $\dot M\approx 10^{-15}M_\odot\,\mathrm{yr}^{-1}$, $B_\mathrm{p}\approx 2\,400$\,G (see sect.\,\ref{planetaryfield}) and $v_\mathrm{w}=300$\,km\,s$^{-1}$ \citep{Preusse2005}, we get $\eta\ast\approx 10^{8}$, meaning that in YZ\,Cet the wind is strongly confined by the magnetic field. Following \cite{Ud-Doula2008}, which give $R_\mathrm{Alf}\approx 0.3+\eta\ast^{1/4}R_\ast$ when $\eta\ast \gg 1$, we get $R_\mathrm{Alf}\approx 100\,R_\ast$, and therefore all the three known planets of YZ\,Cet are inside the inner magnetosphere. In particular, for YZ\,Cet\,b, $v_\mathrm{rel}=85.1$\,km\,s$^{-1}$ (see sect.\,\ref{planetaryfield}), therefore $v_\mathrm{rel}\ll v_\mathrm{w}\ll v_\mathrm{Alf}$ and the planet moves in the Sub-Alfv\'enic region.


\subsection{The hollow cone model}
\label{omega}
Since ARE is highly directive, being the emission pattern either a hollow cone or a narrow beam, as in the case of the tangent beam emission, it is better to visualize the lightcurve in a polar diagram, as shown in Fig.\,\ref{fig:orbit}. Here the visibility of the emission can be correlated with the position of planet b along the orbit. We find that the radio emission is detected only when the planet is in two orbital sectors that are symmetric with respect to the direction of Earth. In the case of the tangent plane beam model, the emission is expected near the quadrature, while here the two sectors are about at $\pm(30^{\circ}$ to $80^{\circ}$) from the direction of the Earth (the two blue sectors in Fig.\,\ref{fig:orbit}). Therefore, our data are consistent with a hollow cone beam model for the ARE.\\
In this model, the emission occurs in the dipolar flux tube connecting the planet and the star, as shown in Fig.\,\ref{fig:dipolo}. The hollow cone has a semi-aperture $\theta$ given by $v/c$, where $v$ is the velocity of the resonant electrons and $c$ the speed of light, and a thickness $\Delta\theta\approx v/c$. The emission is visible from Earth when the line of sight falls inside the walls of the cone. This is shown in the schematic picture of  Fig.\,\ref{fig:cono}, where the hollow cone intercepts the  sphere of unit radius in a circular ring. The points A, B and C, D indicate the moments of start and stop of visibility of ARE before and after the pseudo transit of the planet. Here we assume, for simplicity, that the axis of the dipole of the star coincides with the rotational axis. The visibility depends on $v/c$, on the location of the source of emission in the dipolar loop, which is defined by the angle $\psi$, and the inclination $i$. In order to identify possible values of the parameters, we project the circular emission ring onto the orbital plane, as in Fig.\,\ref{fig:cone-projection}. The circular ring is defined by two ellipses that intercept the projection E$'$ of the line of sight at four points A$'$, B$'$ and C$'$, D$'$.\\
With this simple geometrical model it is possible to infer that $v/c$ lies in the range $0.3-0.8$, and the inclination $i$ is between $30^{\circ}$ and $60^{\circ}$. In Fig.\,\ref{fig:cone-projection} a possible configuration corresponding to the observed emission pattern is shown. The solid angle $\Omega$ subtended by the cone is defined by $\theta$ and $\Delta\theta$, which are given by $v/c$. For the range of $v/c$ that we find, $\Omega\approx 1.8-3$\,sr.

We observe high degree of circular polarization on day 5 and 6, with positive values of Stokes V. This means that, if the emission is in the x-mode, it is produced in the Northern magnetic hemisphere. It is worth noting that the true stellar magnetic field topology and the real geometry of the system (i.e. the dipole axis inclination with respect to the exoplanet orbital plane and respect to the line of sight) are basically unknown. This prevents us from providing firm conclusions regarding some observational evidence, mainly the non-detection of circular polarization on days 3 and 4. We can only suggest that one possible explanation is that on day 3 and 4 we observe radiation emitted from the two hemispheres simultaneously.

\subsection{The Stellar magnetic field}
\label{stellarB}
The spectrum of the ECME is directly connected with the local magnetic field strength $B$. The frequency of the maser is given by $\nu=s\cdot 2.8\,B$\, MHz, where $s=1,2,3,4$ is the harmonic number, $B=B_\mathrm{p}(\frac{R_\ast}{r})^{3}$, $B_\mathrm{p}$ is the magnetic field strength at the pole of the star and $R$ the radius of the magnetosphere above the pole where the ECME forms in a dipolar topology \citep{Trigilio2000}.
We fix $s=2$ as the first harmonic is likely to be suppressed by the second harmonic of the gyrofrequency of the surrounding plasma \citep{MelroseDulk1982, Trigilio2000} and the higher harmonics have a small intensity. 
The spectra in Fig.\,\ref{fig:spettromedio} seem to point to a lower limit cutoff of about $\nu_\mathrm{min}\approx 500$\,MHz, which corresponds to $B_\mathrm{min}\approx90$\,G above the stellar pole, at a distance $r_\mathrm{max}$ from the centre. 
The region of the magnetic loop where the ECME develops can be estimated when the cutoff of the spectrum and the polar magnetic field are known. We have these data for the Jupiter-Io DAM emission and for the mCP star CU\,Vir. For Io-DAM, the spectrum extends from 3 to 30\,MHz \citep{Zarka2004}, with $B_\mathrm{P}=14$\,G. For CU\,Vir, \citet{Das2021} find that the low frequency cutoff is below their observing band, at about 300\,MHz, and the upper frequency cutoff is at about 3000\,MHz, with $B_\mathrm{P}=3000$\,G. For a dipolar field topology, the ECME originates at $r\approx1.4-3\,R_\ast$ from the center of the dipole. Assuming the same range of $r$ for YZ\,Cet, the polar magnetic field strength is
\begin{equation}
B_\mathrm{p}=\frac{\nu_\mathrm{min}/\mathrm{MHz}}{s\cdot 2.8\cdot (R_\ast/r_\mathrm{max})^{3}}\,\mathrm{G}
\end{equation}
that gives $B_\mathrm{p}\lessapprox 2\,400$\,G for $s=2$. This value is in agreement with what is expected for a M4.5V star \citep[e.g.][]{Kochukhov2017,Kochukhov2020} and, in particular, with the value given by \cite{Moutou2017} of 2.2\,kG. This is just a first estimate, as the best value of $B_\mathrm{p}$ can be provided only if the whole spectrum, including the high frequency cutoff, is known. The frequency corresponding to the range of $r$ given above are $\nu_\mathrm{max}-\nu_\mathrm{min}\approx 4900-500$\,MHz, with $\Delta\nu\approx 4400$\,MHz.

\subsection{The Planetary magnetic field}
\label{planetaryfield}

The power emitted by the ECME, inferred from the observations, can be obtained as
\begin{equation}
  \label{power_obs}  P_\mathrm{obs}=F_\mathrm{\nu}\,\Delta\nu\,2\Omega\,d^{2}
\end{equation}
where $F_\mathrm{\nu}=0.5\,$mJy is the average flux density of the emission inside the hollow cone, $\Delta\nu$ is the bandwidth of the ECME (see sect.\,\ref{stellarB}), $\Omega$ the solid angle subtended by the hollow cone of emission (see sect.\,\ref{omega}) and $d$ is the distance to the star. The factor 2 accounts for the two hemispheres that we assume emit the same power but at opposite circular polarization. We get $P_\mathrm{obs}$ in the range between $1.0\times 10^{22}$\,erg\,s$^{-1}$ and $1.7\times 10^{22}$\,erg\,s$^{-1}$, corresponding to the range of $\Omega$. \\
On the other hand, the emitted power $P_\mathrm{obs}$ is a fraction $\epsilon$ of the incident power $P_\mathrm{in}$ due to the interaction between the stellar magnetosphere and the planet \citep[e.g.][]{Zarka2007, Lanza2009}. This is given by:
\begin{equation}
\label{power_in}
P_\mathrm{in}=A\,v_\mathrm{rel}B^2/8\pi
\end{equation}
where $A$ is the cross section of the planet, $v_\mathrm{rel}$ is the relative velocity of the planet to the magnetic field of the star and $B$ the magnetic field of the star at the position $r$ of the planet. Assuming that the orbital plane coincides with the magnetic equatorial plane of the star, $B=\frac{1}{2}\,B_\mathrm{p}(\frac{R_\ast}{r})^{3}$. Since for YZ\,Cet\,b $r=21.9\,R_\ast$, considering that $B_\mathrm{p}$ is an upper limit, $B\leq0.1$\,G. The relative velocity is $v_\mathrm{rel}=\left|v_\mathrm{orb}-v_\mathrm{cor}\right|$, where $v_\mathrm{orb}=87.6$\,km\,s$^{-1}$ is the orbital velocity of the planet and $v_\mathrm{cor}=2.5$\,km\,s$^{-1}$ is the co-rotational velocity at the position of the planet. 
If the planet does not have a magnetic field, $A=\pi\,R_\mathrm{planet}^2$. In this case $P_\mathrm{in}\leq4.3\times 10^{21}$\,erg\,s$^{-1}$, a value that is smaller than $P_\mathrm{obs}$. 
Since $P_\mathrm{in}$ must be $\geq P_\mathrm{obs}$, the only possibility is to consider an increase of the cross section $A$ at least by a factor $2.4$ or $4.0$  corresponding to the range of $\Omega$. This means to consider a planetary magnetic field. In this case $A=\pi\,R_\mathrm{MP}^2$, where $R_\mathrm{MP}$ is the radius the magnetopause, i.e. the distance from the centre of the planet where its magnetic field strength equals $B$. The condition $P_\mathrm{in} \geq P_\mathrm{obs}$ translate in $R_\mathrm{MP} \geq 1.6\, R_\mathrm{Planet}$ and $R_\mathrm{MP} \geq 2.0\, R_\mathrm{Planet}$ corresponding to the range of $\Omega$. Assuming a dipolar field for the planetary magnetosphere:
\begin{equation}
    R_\mathrm{MP}=R_\mathrm{planet}(B_\mathrm{planet}/B)^{1/3}
\end{equation}
the above conditions imply $B_\mathrm{Planet}\geq 0.4$\,G and $B_\mathrm{Planet}\geq 0.9$\,G.
Moreover if the “generalized radio-magnetic Bode's law” with $P_\mathrm{obs} / P_\mathrm{in} \approx 0.01$ \citep{Zarka2007} were valid for our system, the above limits would increase considerably.

\section{Conclusions}
Our main finding is the detection of highly circularly polarized radio emission in the YZ\,Cet system that is consistent with being due to ARE from SPI.
The spectrum of the ARE and the correlation with the position of the planet along the orbit allows us to estimate the magnetic field of the star and the characteristics of the emission cone. 

The comparison between the radiated power and the incident magnetic power allows us to infer the presence of a magnetosphere of the planet.
We estimate a lower limit for the magnetic field of the planet YZ\,Cet\,b of $0.4$\,G. 
If confirmed, this would be the first (indirect) measurement of a planetary magnetic field. 

We find that the strong radio emission from the interaction between YZ\,Cet\,b and its star is detected only when the planet is in two orbital sectors that are symmetric with respect to the direction of Earth. This behavior can be explained within the framework of the hollow cone beam model for the ARE, and is in contrast with the tangent plane beam, where the emission is expected to increase near the quadratures, as found in e.g. the Proxima~b - Proxima system \citep{Perez-Torres2021}. 

Radio follow-up observations of this system will allow us to constrain better some of the most relevant parameters responsible for the observed ARE, e.g., the low- and high-frequency cutoffs, or the solid angle, $\Omega$ covered by the ARE. 

We emphasize that our work outlines a promising method for the study of SPI and for the indirect detection of planetary magnetospheres. Both are important in defining the exoplanet environment and hence the possibility of favourable conditions for the evolution of life.

\begin{acknowledgments}
We thank the staff of the GMRT who have made these observations possible. The GMRT is run by the National Centre for Radio Astrophysics of the Tata Institute of Fundamental Research. We acknowledge support of the Department of Atomic Energy, Government of India, under project no. 12-R\&D-TFR-5.02-0700.
BD acknowledges support from the Bartol Research Institute.
MPT acknowledges financial support through grants CEX2021-001131-S and PID2020-117404GB-C21 funded by the Spanish MCIN/AEI/ 10.13039/501100011033.
\end{acknowledgments}

\bibliography{YZCet}{}
\bibliographystyle{aasjournal}



\end{document}